\documentclass[twocolumn]{aastex6}
\usepackage{amsmath}
\usepackage{amssymb}
\usepackage{latexsym}
\usepackage{listings}
\usepackage{graphicx}
\usepackage{rotating}
\usepackage{url}
\usepackage{theorem}
\usepackage{verbatim}
\usepackage{natbib}
\DeclareGraphicsExtensions{.pdf}
\shorttitle{Comprehensive Analysis of GJ 1243}
\shortauthors{Silverberg et al.}

\begin{document}

\title{Kepler Flares IV: A Comprehensive Analysis of the Activity of the \lowercase{d}M4\lowercase{e} Star GJ 1243}
\author{Steven M. Silverberg\altaffilmark{1},
Adam F. Kowalski\altaffilmark{2,3},
James R. A. Davenport\altaffilmark{4,5},
John P. Wisniewski\altaffilmark{1},
Suzanne L. Hawley\altaffilmark{6},
Eric J. Hilton\altaffilmark{6}
}

\altaffiltext{1}{Homer L. Dodge Department of Physics and Astronomy, University of Oklahoma, 440 W. Brooks Street, Norman, OK 73019,USA}
\altaffiltext{2}{Department of Astronomy, University of Maryland, College Park, MD 20742, USA}
\altaffiltext{3}{NASA Goddard Space Flight Center, Code 671, Greenbelt, MD 20771, USA}
\altaffiltext{4}{Western Washington University, Bellingham, WA 98225}
\altaffiltext{5}{NSF Astronomy and Astrophysics Postdoctoral Fellow}
\altaffiltext{6}{Department of Astronomy, University of Washington, Box 351580, Seattle, WA 98195, USA}

\begin{abstract}
We present a comprehensive study of the active dM4e star GJ 1243. We use previous observations and ground-based echelle spectroscopy to determine that GJ 1243 is a member of the Argus association of field stars, suggesting it is $\sim 30-50$ Myr old. We analyze eleven months of 1-minute cadence data from \textit{Kepler}, presenting \textit{Kepler} flare frequency distributions, as well as determining correlations between flare energy, amplitude, duration, and decay time. We find that the exponent $\alpha$ of the power-law flare energy distribution varies in time, primarily due to completeness of sample and the low frequency of high-energy flares. We also find a deviation from a single power law at high energy. We use ground-based spectroscopic observations simultaneous with the \textit{Kepler} data to provide simultaneous photometric and spectroscopic analysis of three low-energy flares, the lowest-energy dMe flares with detailed spectral analysis to date on any star. The spectroscopic data from these flares extend constraints for radiative hydrodynamic (RHD) flare models to a lower energy regime than has previously been studied. We use this simultaneous spectroscopy and \textit{Kepler} photometry to develop approximate conversions from the \textit{Kepler} bandpass to the traditional $U$ and $B$ bands. This conversion will be a critical factor in comparing any \textit{Kepler} flare analyses to the canon of previous ground-based flare studies.
\end{abstract}

\section{Introduction}
M dwarfs are well known for their high magnetic activity, most notably their powerful, frequent flares. These are believed to be magnetic reconnection events analogous to the flares observed on the Sun, but occur much more frequently, and with much greater energies. Observational studies of flares on these stars have been limited by the difficulty of collecting statistically complete samples of detailed flare light curves from ground-based observation. To mitigate the stochastic occurence rate of flares, studies have often chosen to follow multiple active stars with similar spectral types over several nights \citep[e.g.][]{Moffett1974, Hilton2011}. Automated surveys provide repeated imaging of the sky and can efficiently yield millions of individual measurements of M dwarfs, but they do not provide temporal information for individual flares, or observe a complete sample of flares for individual stars. 

Dedicated ground- and space-based exoplanet surveys, however, provide sufficiently high-cadence observations over a duration long enough to ensure a statistically complete sample of flares on observed stars. Due to the inclusion of local M dwarfs in transiting exoplanet surveys such as MEarth \citep{Nutzman2008} and \textit{Kepler} \citep{Borucki2010}, detailed study of M dwarf activity in the solar neighborhood has dramatically increased in recent years. The \textit{Kepler} survey launched a new era of stellar photometric investigation, allowing for unprecedented light curve collection. While the mission's primary goal has been the detection of exoplanets, its near-continuous short- and long-cadence monitoring of targets make it nearly ideal for recording statistically complete samples of stellar variability, including flares, throughout the main sequence \citep{Basri2010, Walkowicz2011}.

Previous works  have utilized \textit{Kepler}'s unique capabilities to explore M dwarf activity with unparalled depth and completeness. \citet[hereafter Paper 1]{Hawley2014} analyzed the flare frequency distribution of three active and three inactive M dwarfs with two months of high-cadence \textit{Kepler} data, finding strong correlations between flare energy, amplitude, duration, and decay time, and a weak correlation with rise time. No correlation was found between flare energy or occurrence and starspot phase, and energies of consecutive flares. \citet[hereafter Paper 2]{Davenport2014} expanded on the sample from Paper 1, producing a 90\% complete sample of over 6100 flares from eleven months of \textit{Kepler} short-cadence data collected on GJ 1243. \citet[hereafter Paper 3]{Lurie2015} used \textit{Kepler} data on the active stars GJ 1245 A and B to study the evolution of flares and starspots in a binary system, providing a detailed photometric analysis of a multiple system of fully convective M dwarfs and yielding an important constraint on stellar age-rotation-activity models. Paper 3 also found that some flares for GJ 1245 were bright enough that their peak fluxes equaled or exceeded 95\% of \textit{Kepler}'s quoted full well depth, potentially entering the non-linear regime and saturating the CCD. This paper, the fourth in this series, focuses on the properties of GJ 1243.

GJ 1243 is a dMe flare star located in the solar neighborhood \citep[see e.g.][]{Reid2004}. \citet{Hawley1995} and \citet{Hawley1996} identified the star as a dMe based on strong H$\alpha$ emission (EW $>$ 1 \AA) as part of a survey of the preliminary version of the Third Catalog of Nearby Stars \citep{GJ1991}, while \citet{Gershberg1999} later identified the star as a a UV Ceti-type flare star, and \citet{Gizis2002} confirmed it as active. The MEarth survey found a distance to the star of $13.48 \pm 0.42$ pc based on trigonometric parallax \citep{Dit2014}. Additionally, \citet{LepineShara2005} found that GJ 1243 was a high-proper motion star, with a root mean square proper motion of 326 $\mathrm{mas\,yr^{-1}}$. More recently, \citet{Irwin2011} found a period for GJ 1243 of 0.593 days using MEarth photometry. \citet{Savanov2011} used long-cadence data from \textit{Kepler} Q1 to confirm a period of 0.5926 days, and identified two independent regions of starspots that persisted in the same location throughout the observation. \citet{Reinhold2013} found no evidence of differential rotation in GJ 1243 as part of a larger survey of \textit{Kepler} targets, but \citet{Davenport2015} found evidence of differential rotation using phase-tracking and spot modeling. \citet{Ramsay2013} also noted frequent flares on GJ 1243 as a comparison to KIC 5474065 (also an M4V star), and conducted a flare rate analysis based on short-cadence data from \textit{Kepler} Q6 (also used in Paper 1).

In this paper, we combine the techniques of all previous work in the \textit{Kepler} Flares series to provide a comprehensive analysis of the behavior of GJ 1243 for the full eleven months of short cadence data collected. In Section 2, we review the methods of observing and recording both photometric and spectroscopic data, and discuss how flares are identified in both data sets. In Section 3, we discuss the kinematics and intrinsic characteristics of GJ 1243: its rotational velocity, proper motion in the galaxy, magnetic field, and metallicity. In Section 4, we analyze the full flare sample from \textit{Kepler}, investigating the flare frequency-energy relationship and the correlations found in Paper 1. We also examine these in shorter (month-by-month and quarter-by-quarter) time units, searching for predictable changes over time. In Section 5, we examine spectroscopic observations of GJ 1243 collected simultaneously with \textit{Kepler} observations to break modeling degeneracies that result from the single bandpass on \textit{Kepler}, and characterize the spectral behavior of the observed flares. Finally, in Section 6, we discuss these observations and build a full characterization of GJ 1243.

\section{Observations and Data}

\subsection{Flare Photometry from \textit{Kepler}}

As the most active M dwarf in the \textit{Kepler} field, GJ 1243 was the primary subject of multiple \textit{Kepler} ``Guest Observer'' campaigns (GO programs 20016, 20028, 20031, 30002, 30021). These campaigns yielded 11 months of short cadence data from the primary \textit{Kepler} mission. Paper 2 used what was at the time the most recent reduction of the \textit{Kepler} light curves, including the PDC-MAP Bayesian detrending analysis from \citet{Smith2012}, to build its flare sample, which we use here.

The flare identification process we adopt here was thoroughly detailed in Paper 2. Briefly, flare candidates were first identified by automatically selecting events of two or more time steps in length with positive flux excursions greater than 2.5$\sigma$ from the starspot-subtracted light curve. After this, all data were visually inspected using the FBEYE package\footnote{Available online at \url{https://github.com/jradavenport/FBEYE}} to validate the initial computer classifications. Multiple users classified each month of data, and the final flare sample\footnote{\noindent Available online at \url{http://github.com/jradavenport/GJ1243-Flares}} was selected from a composite of all user flare identifications. Flare start and end times were selected to include all observations identified as part of a flare by at least two users.

We note that since the publication of Paper 2, additional errors in the short cadence data processing have been uncovered, which impact the data on GJ 1243 from \textit{Kepler} Q10 and Q12.\footnote{For more information see this erratum: \url{http://keplerscience.arc.nasa.gov/data/documentation/KSCI-19080-001.pdf}} The amplitude of these calibration errors is typically small. However, as the impact on this specific data set is as yet unknown, we note that some caution must be taken when interpreting the rates of the smallest energy flares. Future versions of this work will utilize Data Release 25 when it becomes available in mid- to late-2016.

\subsection{Flare Spectroscopy from DIS}

To break the modeling degeneracy due to \textit{Kepler}'s single bandpass, we obtained ground-based spectroscopic observations of the star simultaneous with \textit{Kepler} observations. Spectra for GJ 1243 were collected with the Dual-Imaging Spectrograph (DIS) on the Astrophysical Research Consortium (ARC) 3.5 m telescope at the Apache Point Observatory (APO) on 2012 May 18 (during \textit{Kepler} Q13), following the procedure described in \citet{Kowalski2013}. The low-resolution B400/R300 gratings were used, providing continuous wavelength coverage from $\lambda \sim 3400-9200$ \AA, except for a dichroic feature that affected flux calibration at $\lambda \sim 5200-5900$ \AA. Exposure times ranged from 45 to 60 s, with $\sim 10$ s readout times. Short cadence spectra were interspersed to avoid non-linearity and saturation in the red; this typically provided a signal-to-noise of $\sim 10$ at 3600 \AA. GJ 1243 was observed with the 5'' slit, which facilitates absolute flux calibration, mitigates the effects of slit loss, and allows for reduced exposure times. The slit was oriented perpendicular to the horizon to compensate for atmospheric differential refraction.

The spectra were reduced using standard IRAF procedures via a customized PyRAF wrapper, developed from the reduction software of \citet{Covey2008}. Wavelengths were calibrated from HeNeArHg and HeNeAr lamps, resulting in dispersions of 1.83 \AA\ pixel$^{-1}$ in the blue and 2.3 \AA\ pixel$^{-1}$ in the red. Spectral resolutions were determined from the He I $\lambda4471$ arc line recorded at the start of observations, resulting in a resolution of $\sim 18$ \AA\ ($R \sim 250$); however, for the 5'' slit, the spetral resolution was determined by the seeing, leading to a higher resolution on the observed objects ($R\sim 320$). To correct for the radial velocity uncertainty that results from the wide slit, we found it necessary to apply a wavelength shift to the blue and red spectra using the centers of the H$\gamma$ and H$\alpha$ emission lines, respectively. The spectrophotometric standard star BD+28 4211 was observed and used to flux-calibrate the spectra, and an airmass correction was applied using the atmospheric extinction curve for APO published by the Sloan Digital Sky Survey (SDSS). Despite this airmass correction, we observe a residual $5-8\%$ color-independent variation in spectrum brightness, which corresponds well to the change in airmass; spectra generally get brighter as airmass decreases through the night. With multiple standard star observations through the night, a residual airmass correction could be applied; however, to maximize the number of observations simultaneous with \textit{Kepler}, we only observed a standard for flux-calibration purposes at the start of the night (at airmass 1.5).

\subsection{Echelle Spectroscopy}

We collected optical spectra of  GJ 1243 and Gliese 406 (Wolf 359) on UT 23 June, 2011, with the Astrophysical Research Consortium Echelle Spectrograph (ARCES) on the ARC 3.5m telescope at APO. ARCES \citep{Wang2003} is a high-resolution, cross-dispersed spectrograph, collecting $R \sim 31,500$ spectra between 3600 and 10000 \AA. In addition to these stars, we recorded bias, flat-field, and ThAr lamp exposures. These data were then reduced with standard IRAF procedures.\footnote{IRAF is distributed by the National Optical Astronomy Observatories, which are operated by the Association of Universities for Research in Astronomy, Inc., under cooperative agreement with the National Science Foundation.}

\section{Kinematics and Intrinsic Characteristics}
\label{section:kinematics}

\begin{figure}[tb]
\begin{centering}
\includegraphics[width=.5\textwidth]{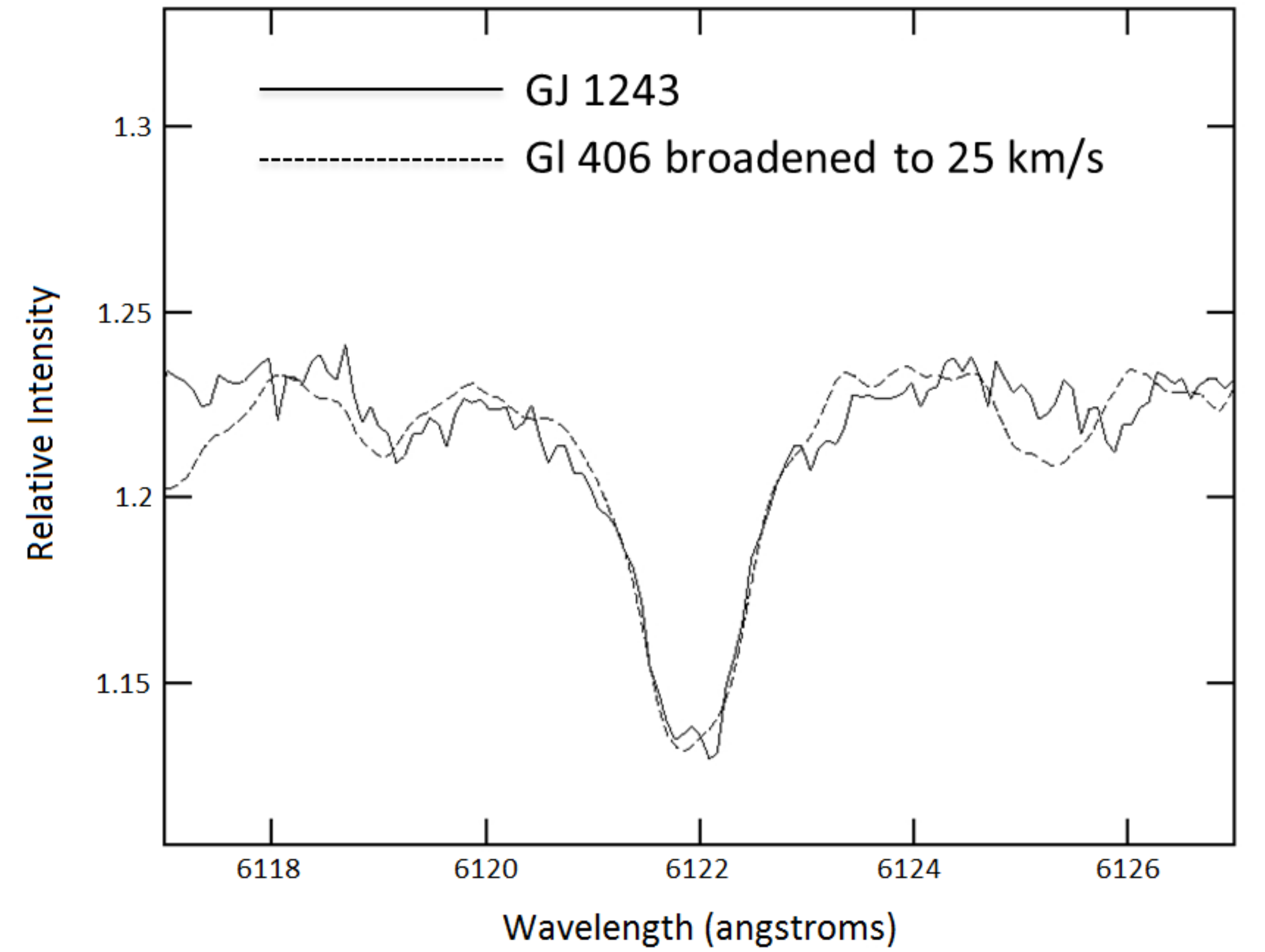}
\caption{Echelle observations ($R = 30,000$) of the Ca I line at 6122 \AA\ on GJ 1243 and Gliese 406 (Wolf 359) from the ARC 3.5m telesope. An artificial rotational broadening of $25$ km/s has been applied to the Gliese 406 spectrum, which matches quite well with the line profile of GJ 1243.}  
\label{fig:vsinifig}
\end{centering}
\end{figure}

To determine $v \sin i$, we analyzed our echelle observations of GJ 1243 and Gliese 406. The spectrum for Gl 406 was artificially rotationally broadened until the profile of the Ca I line at 6122 \AA\ in Gl 406 matched that of GJ 1243. Figure \ref{fig:vsinifig} shows that $v \sin i \sim 25 \, \mathrm{km\, s^{-1}}$ provides a good qualitative match to GJ 1243. Paper 1 found that GJ 1243 has spectral type M4, which has a representative radius of 0.36 $R_{\odot}$ \citep[e.g.][]{Hawley1995}. Based on this assumed radius and the 0.59-day period observed in ground-based and \textit{Kepler} long-cadence observations, the inclination angle is $\sim 32$ degrees. This indicates that some circumpolar areas of the star are visible at all times, a relevant factor in understanding the phase of the star, as well as inferring the distribution of activity across the stellar surface.

\begin{figure}[tb]
\begin{centering}
\includegraphics[width=.5\textwidth]{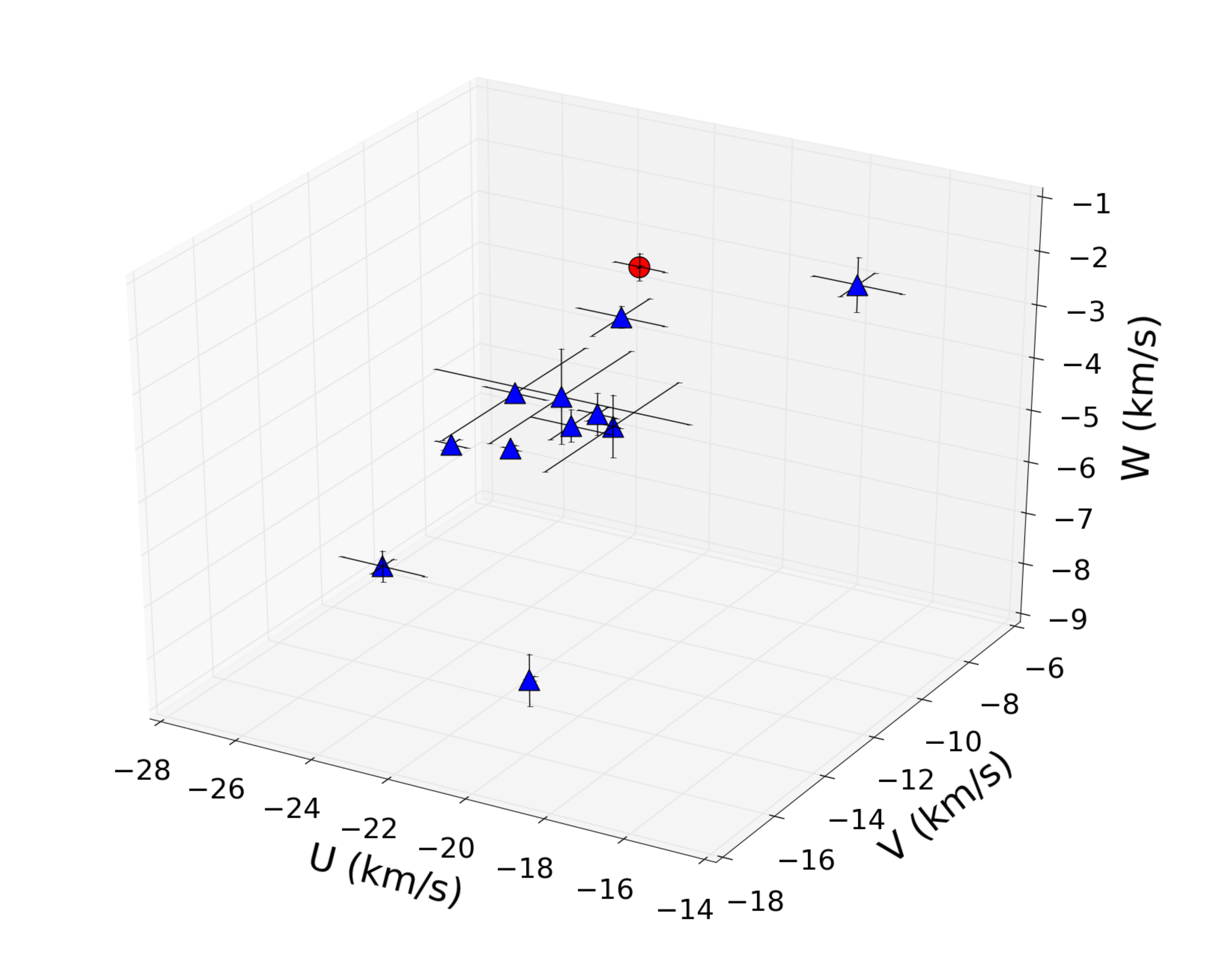}
\caption{A three-dimensional projection of GJ 1243 (red circle) and the 11 members of the Argus association (blue triangles) used in BANYAN II, in Galactic velocity space. ($U$ is positive toward the Galactic center.)} 
\label{fig:Argus_velocity_space}
\end{centering}
\end{figure}

Using right ascension and declination proper motions of $188\, \mathrm{mas\, yr^{-1}}$ and $266\, \mathrm{mas\, yr^{-1}}$ \citep{LepineShara2005} respectively, the MEarth parallax of $74.20 \pm 2.30\, \textrm{mas}$ \citep{Dit2014}, and a radial velocity of $-11.38 \pm 0.19\, \mathrm{km\, s^{-1}}$ \citep{Deshpande2013}, the online Bayesian analysis tool BANYAN II \citep{Malo2013, Gagne2014} yields a 98.50\% likelihood that GJ 1243 is a member of the Argus association, with a 1.50\% chance of it being a field star. The Argus association was first identified in the SACY survey \citep{Torres2003b} due to its unique velocity properties in that survey; as seen in Figure \ref{fig:Argus_velocity_space}, which shows the Galactic velocities of GJ 1243 and the Argus association members, it has a relatively large velocity away from the Galactic center. It was more formally defined in \citet{Torres2008}. Its age is estimated to be 30-50 Myr, suggesting that GJ 1243 is quite young relative to other nearby M dwarfs.


\section{Behavior of Flares in White Light}

The final sample of flares from Paper 2, analyzed here, contains 6107 unique flare events, of which 945 (15.5\%) were labeled ``complex'' (e.g. multi-peaked). This sample is the largest catalog of stellar flares from a single object (excluding the Sun) known to the authors. Here, we discuss the general properties of this sample: the cumulative flare frequency diagram (FFD) as a function of time; correlations between flare energies, durations, and amplitudes; and the long-term behavior of flare timing with respect to the observed stellar phase.

\subsection{Flare Frequency}

\begin{figure}[tb]
\begin{centering}
\includegraphics[width=.5\textwidth]{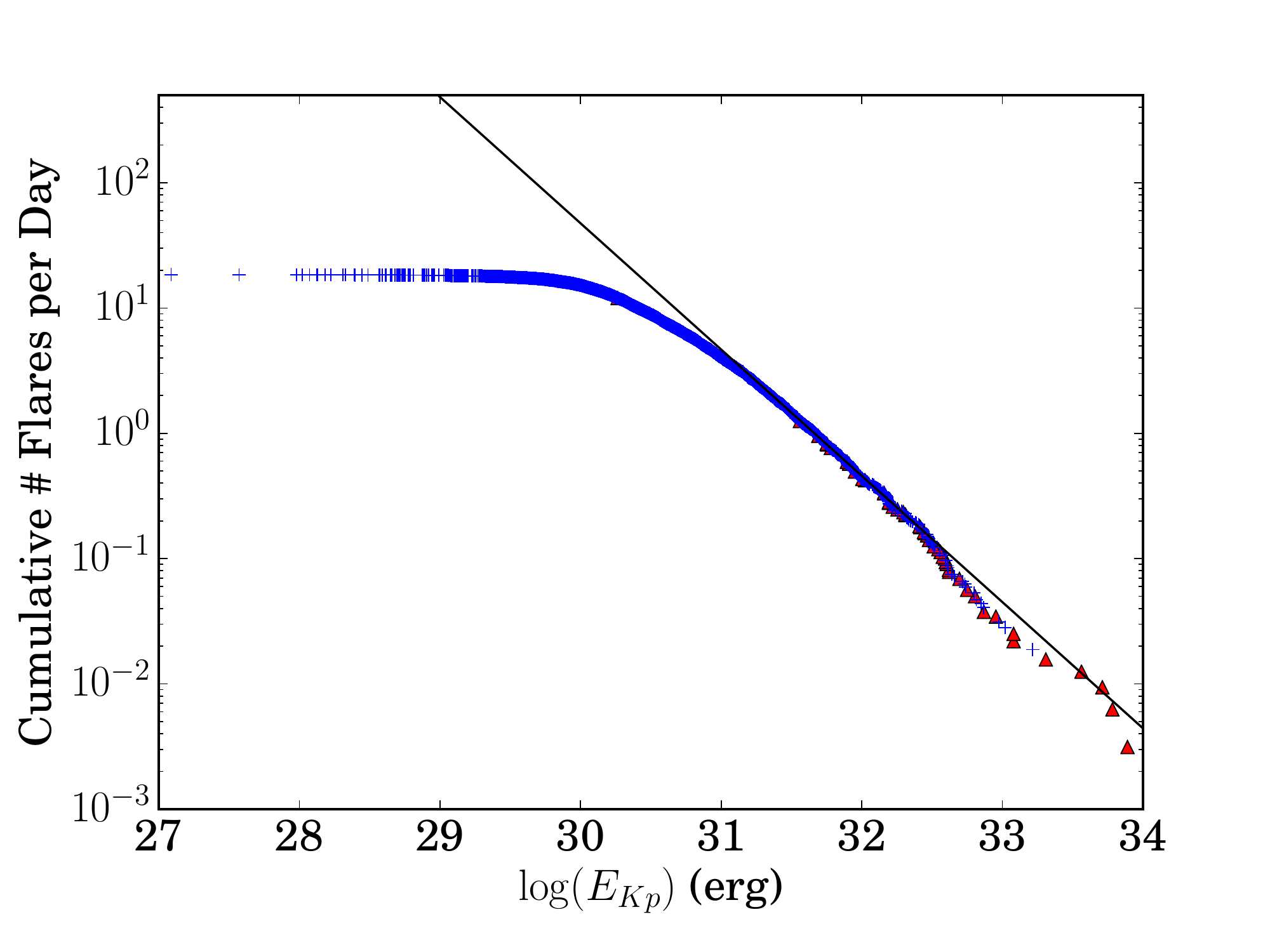}
\caption{The cumulative flare frequency distribution, with least-squares power-law fit, for the active M star GJ 1243 in the \textit{\textit{Kepler}} sample, using flare energies in the \textit{Kepler} bandpass. Red triangles indicate flares where peak fluxes are 95\% of the full well depth or above. The fit lines for a step size for increasing $E_{\mathrm{lim}}$ of 0.05 is plotted in black; other step sizes are nearly identical.} 
\label{fig:All_data_FFD}
\end{centering}
\end{figure}

\begin{figure}[htb]
\begin{centering}
\includegraphics[width=.5\textwidth]{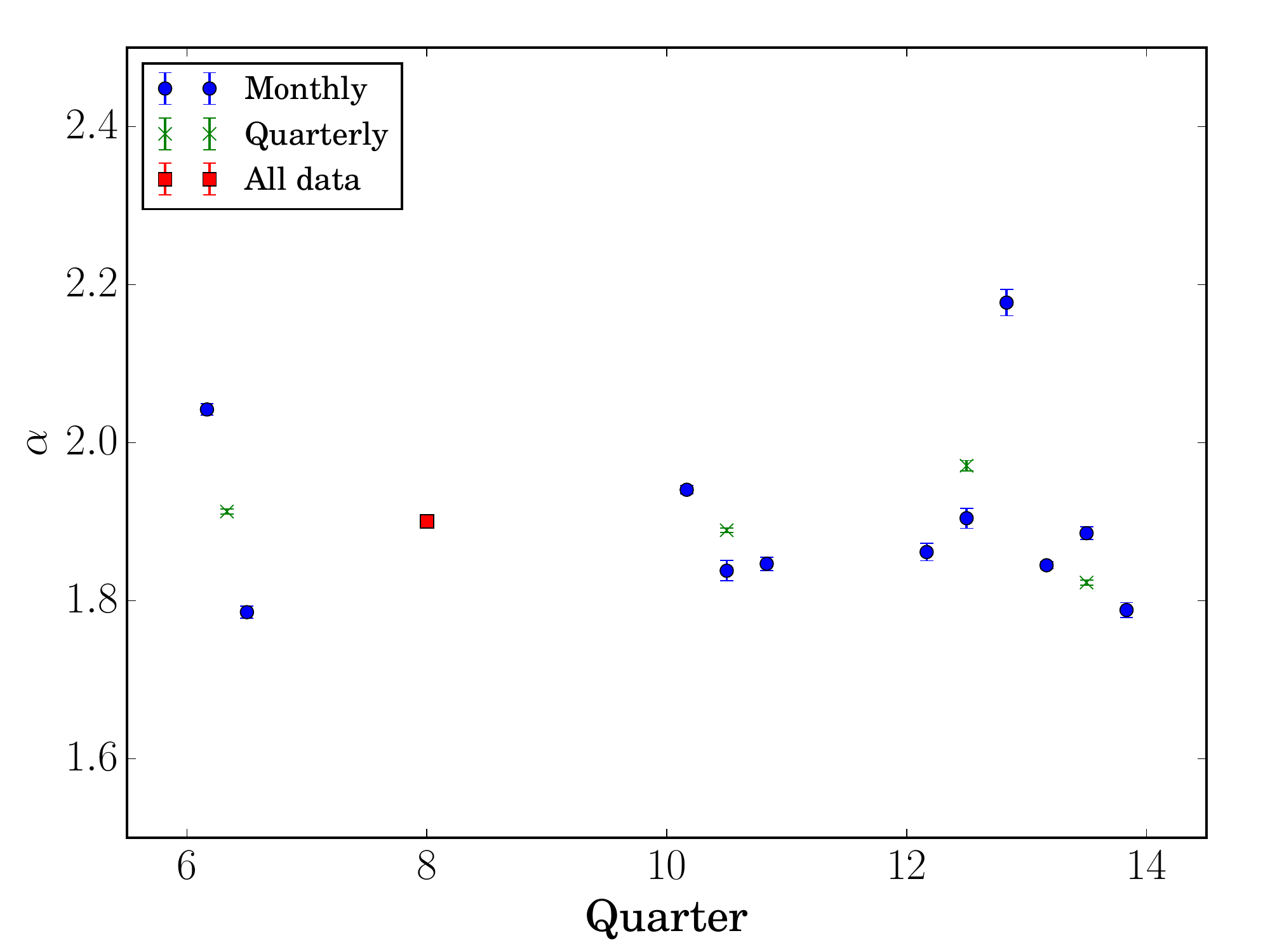}
\caption{Derived values of $\alpha$ for each segment of time in the \textit{\textit{Kepler}} sample, as well as the full data set, as  a function of time, when adopting a fixed energy cutoff at $\log E = 31$. While there is apparent} variability month-to-month and quarter-to-quarter, there does not appear to be any coherent evolution in time. 
\label{fig:alph_v_time_fixed}
\end{centering}
\end{figure}

\begin{figure}[htb]
\begin{centering}
\includegraphics[width=.5\textwidth]{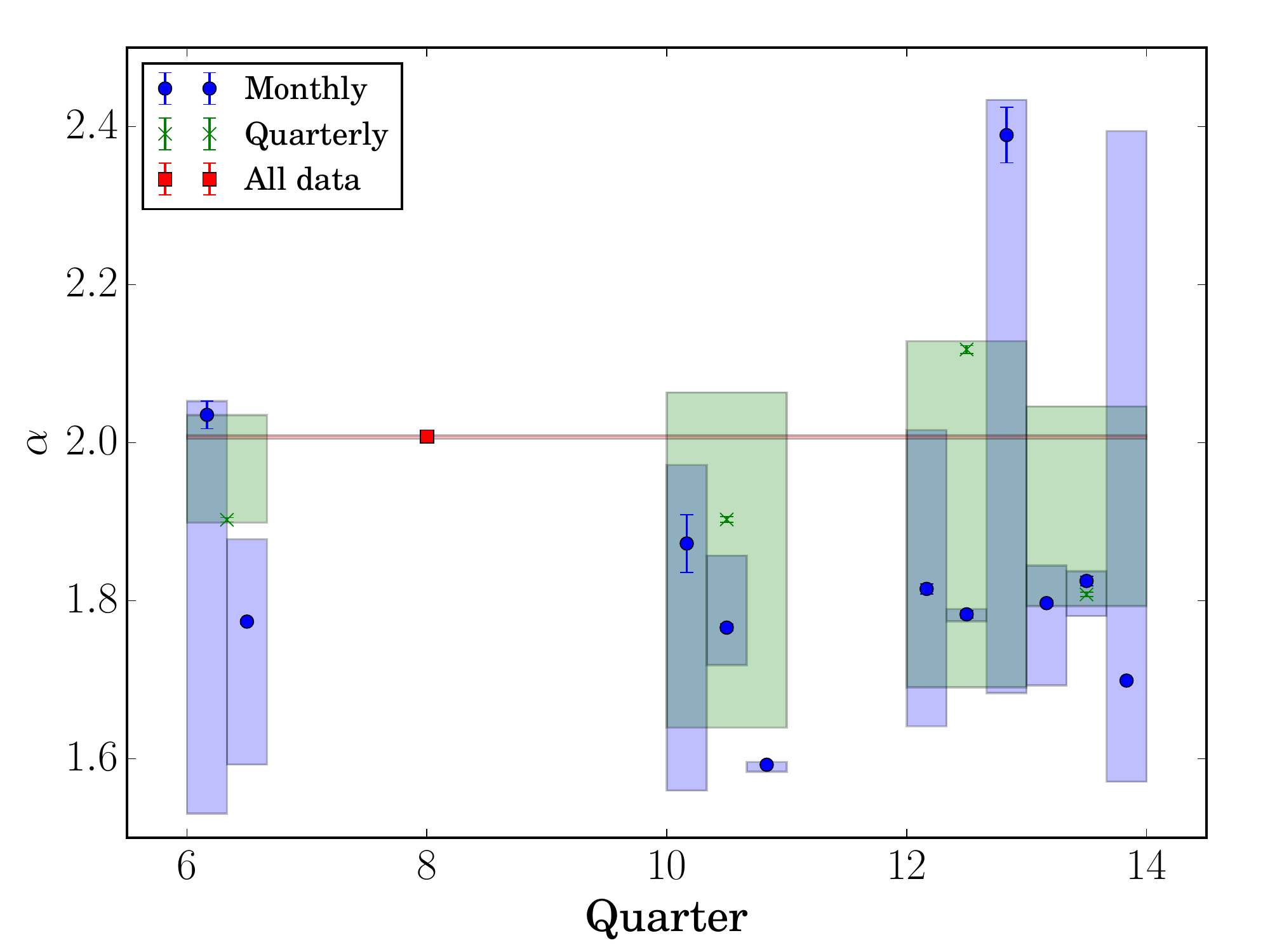}
\caption{Derived values of $\alpha$ for each segment of time in the \textit{\textit{Kepler}} sample, as well as the full data set, as  a function of time. Shaded regions indicate the possible values of $\alpha$ for each segment in time, depending on choice of step size when increasing $E_{\mathrm{lim}}$.} 
\label{fig:alph_v_time}
\end{centering}
\end{figure}

Flare frequency diagrams (FFDs), plots of the cumulative frequency of flares (log number of flares per day with energy greater than $E$), as a function of the energy (log $E$), were constructed for the entire data set and for each month of data, as well as quarterly; the FFD for the full data set is seen in Figure \ref{fig:All_data_FFD}. Following common practice, we characterize the probability distribution for flare energies as a power law, 

\begin{equation}
	N(E) dE \propto E^{-\alpha}dE,
\end{equation}

\noindent where $1 - \alpha$ is the slope of a linear fit to a log-log representation of the cumulative FFD. We characterize the cumulative FFD as a power law for energies above a limiting energy $E_{\mathrm{lim}}$ determined primarily by the completeness of the sample for each segment in time. The power-law fits we obtain for our FFDs of GJ 1243 were determined with a least-squares fit to the flares above $E_{\mathrm{lim}}$. To mitigate the potential over-weighting of the highest-energy flares, where flare energies are much more uncertain (see Paper 3), Poisson uncertainties were used to weight the fit toward energy ranges with better sampling. We did not exclude flares for which peak fluxes are 95\% or more of the quoted full depth of the chip, due to the lack of available data; using these flares with the calculated energies as lower limits still provides more information than excluding them entirely.  We initially fit the data using the energy cutoff of $\log(E_{\mathrm{lim}}) = 31$ found in Paper 1; values of $\alpha$ over time based on this cutoff are found in Figure \ref{fig:alph_v_time_fixed}. We also evaluated $\alpha$ by determining a cutoff for each data subset (monthly, quarterly, and the full set). To do this, we increased $E_{\mathrm{lim}}$ and re-fit until the change in slope did not exceed the uncertainty of the fit. To counter-act potential biases in the determination of $E_{\mathrm{lim}}$ (and thus $\alpha$) due to the step size for the increase in $E_{\mathrm{lim}}$, we ran fits for a variety of step sizes, ranging from $\triangle \log(E_{\mathrm{lim}}) = 0.025$ to $0.125$. The values of $\alpha$ found for each step size often had comparable $\chi^{2}$ values, but the actual values could vary widely. Our adopted value for $\alpha$ in each segment of time (month or quarter) is instead based on a visual inspection of these final values, determining whether the fit line fit the distribution of flares adequately by eye, rather than being confounded by an additional outlier. The range of possible values of $\alpha$ depending on the choice of step size is represented by the shaded regions in Figure \ref{fig:alph_v_time}.

Our values for $\alpha$ and limiting energy $E_{\mathrm{lim}}$ for each segment in time are listed in Table \ref{table:alphas}. We note that for Q6, we adopt the flare identifications from Paper 2, rather than the initial flare identifications made in Paper 1. This leads to a small difference between our reported value of $\alpha$ for this period and those reported in Paper 1.

There is apparent variation in $\alpha$ over time. As seen in Figure \ref{fig:alph_v_time_fixed}, there is no clear pattern to this variation, even when holding a fixed energy cutoff of $\log(E_{\mathrm{lim}}) = 31$ (the cutoff found in Paper 1). This suggests that the variation is likely due to the varying levels of completeness for each sample. Variation in time is more apparent in the adopted values for $\alpha$ when determining the cutoff via the iterative method described above, as seen in Figure \ref{fig:alph_v_time}. However, as the large range of the shaded regions in Figure \ref{fig:alph_v_time} suggests, systematic error from the choice of step size is a large factor here, and thus makes the variation less apparent overall. Determining $\log(E_{\mathrm{lim}})$ separately for each epoch allows each fit to probe to a much higher $E_{\mathrm{lim}}$ in the sample; this effect, combined with the larger sample size, is why $\alpha$ for the full data set in Figure \ref{fig:alph_v_time} is higher than the mean of $\alpha$ for the individual epochs. The systematic error due to step size choice, however, makes this method of evaluation less effective for evaluating time evolution on monthly and quarterly timescales. The large systematic error is generally consistent with the conclusion that the completeness of each sample is a key factor; the systematic error is much lower for the full ($\sim90\%$ complete) data set than it is for each time segment. We discuss further evidence to support this conclusion in Section \ref{sec:correlations}.

\begin{deluxetable}{lcccc}
\tablecaption{Power-law Fits to \textit{Kepler} FFDs for GJ 1243
\label{table:alphas}}
\tablehead{\colhead{Month} & \colhead{Quarter} & \colhead{$\alpha$} & \colhead{$\log(E_{\mathrm{lim}})$} & \colhead{$\alpha(\log(E_{\mathrm{lim}}) = 31)$} }
\startdata
1 & 6a & $2.035 \pm 0.018$ & 31.3 & $2.042 \pm 0.007$ \\
2 & 6b & $1.774 \pm 0.003$ & 30.6 & $1.785 \pm 0.008$ \\
3 & 10a & $1.872 \pm 0.037$ & 31.7 & $1.940 \pm 0.005$ \\
4 & 10b & $1.766 \pm 0.005$ & 30.6 & $1.838 \pm 0.013$ \\
5 & 10c & $1.592 \pm 0.003$ & 30.3 & $1.847 \pm 0.008$ \\
6 & 12a & $1.815 \pm 0.006$ & 30.8 & $1.862 \pm 0.011$ \\
7 & 12b & $1.783 \pm 0.005$ & 30.7 & $1.904 \pm 0.013$ \\
8 & 12c & $2.389 \pm 0.035$ & 31.3 & $2.177 \pm 0.017$ \\
9 & 13a & $1.797 \pm 0.004$ & 30.8 & $1.845 \pm 0.004$ \\
10 & 13b & $1.825 \pm 0.006$ & 30.9 & $1.885 \pm 0.008$\\
11 & 13c & $1.699 \pm 0.004$ & 30.6 & $1.788 \pm 0.009$\\
\cline{1-5}
\\
$1-2$ & 6 & $1.902 \pm 0.003$ &  31.0 & $1.913 \pm 0.003$ \\
$3-5$ & 10 & $1.903 \pm 0.004$ & 31.1 & $1.889 \pm 0.003$ \\
$6-8$ & 12 & $2.118 \pm 0.005$ & 31.3 & $1.971 \pm 0.007$ \\
$9-11$ & 13 & $1.808 \pm 0.003$ & 31.0 & $1.823 \pm 0.003$ \\
\cline{1-5}
\\
All data & $-$ & $2.008 \pm 0.002$ & 31.5 & $1.901 \pm 0.001$\\
\enddata
\end{deluxetable}

It is particularly interesting to examine the behavior of the highest energy flares, which show clear variation around the single-power-law fit and may contribute to the large range of values of $\alpha$ seen in certain time segments (months 1, 3, 8, and 11). \citet{Maehara2015} searched for a separate power law for the highest-energy flares on solar-type stars, with a fixed bolometric energy lower limit of $\log(E) = 32$. However, for several reasons, this cutoff cannot simply be applied directly. The \citet{Maehara2015} energies are the inferred bolometric flare energies, rather than specifically those in the \textit{Kepler} bandpass, making the two measurements incompatible. Additionally, a solar-type star can achieve a higher energy flare at lower relative amplitude, due to its higher intrinsic luminosity (using the equivalent duration method of calculating energy); this leads to far fewer saturated flares at these energies on solar-type stars than it does in our data set. As seen in Figure \ref{fig:All_data_FFD}, the majority $(\sim 78 \%)$ of the 49 flares in which the peak fluxes equaled or exceeded 95\% of \textit{Kepler}'s quoted full well depth occur above the $\log(E_{\textit{Kp}}) = 32$ limit, but these saturated flares also occur down to energies of $\log(E_{\textit{Kp}}) \sim 31$.

\begin{figure}[htb]
\begin{centering}
\includegraphics[width=.5\textwidth]{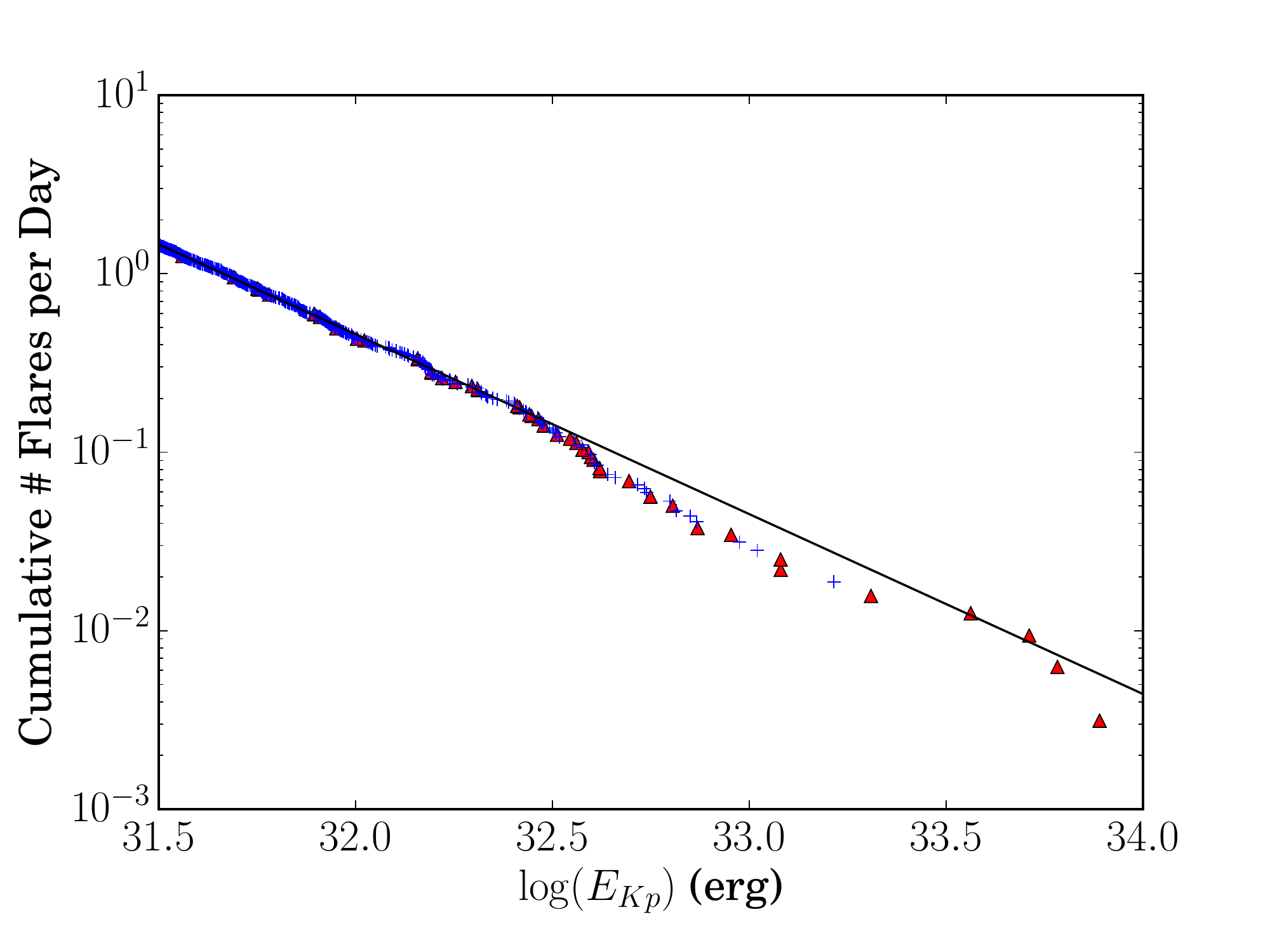}
\caption{An enlarged view of the FFD for GJ 1243, focusing on high energies. Both saturated and unsaturated flares deviate from the power law at approximately $\log(E_{\textit{Kp}}) = 32.5$, suggesting that the deviation observed in Paper 3 is not exclusively due to saturation effects.}
\label{fig:Paper_fig_new_zoom}
\end{centering}
\end{figure}

Paper 3 found a deviation from a single power-law fit at high energies, but attributed this exclusively to the lower-limit effect of the nonlinearity of some high-energy flares. However, close inspection of the FFD at high energies (as in Figure \ref{fig:Paper_fig_new_zoom}) shows that this deviation occurs in \textit{un}saturated flares of high energy. This suggests that the break-away from the single power law seen in the FFDs here and in Paper 3 is not exclusively a factor of entering the saturated regime of the CCD, which could correspond to a different power law at high energies. The relatively small number of flares in question suggests that this difference could be due simply to small-number statistics. Nevertheless, a significant change in slope from the single-power-law model has many implications for our understanding of flare generation and behavior, and warrants further study. Future work will include the addition of high-energy flares found in \textit{Kepler} long-cadence data, to improve the completeness of the sample at these energies.

\subsection{Flare Amplitude, Duration, and Energy}
\label{sec:correlations}

\begin{figure*}[htb]
\begin{centering}
\includegraphics[width=\textwidth]{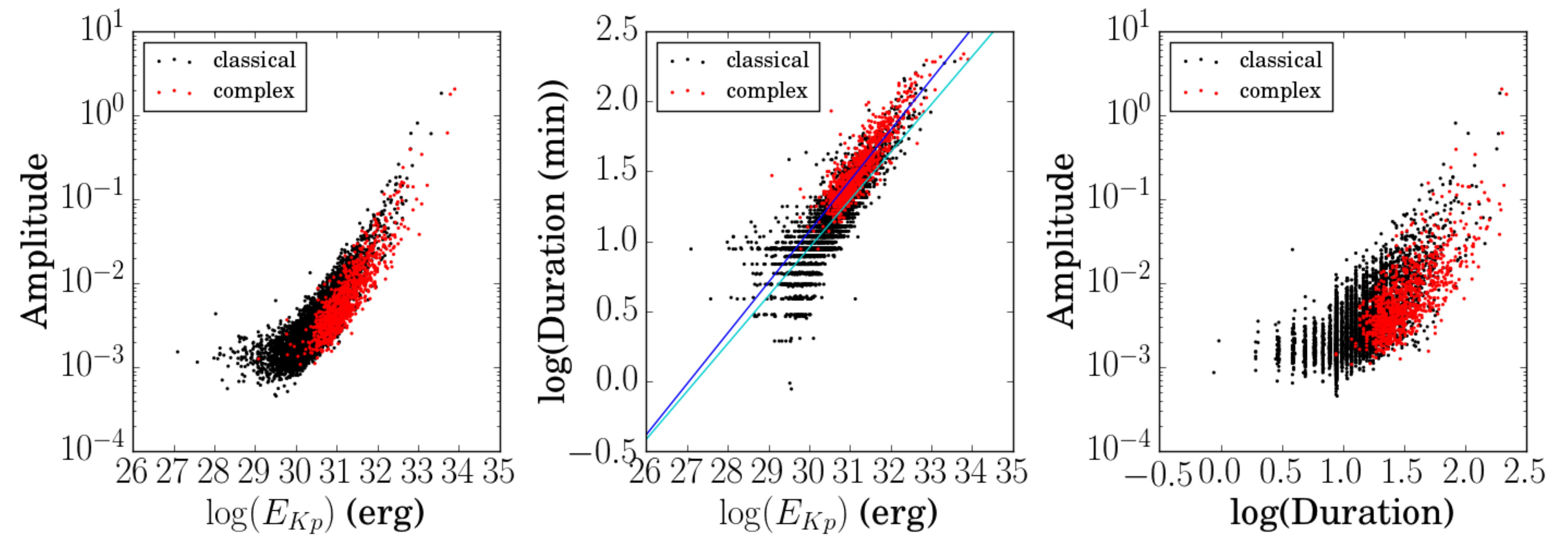}
\caption{Relationships between amplitude, energy, and duration for both classical and complex flares on GJ 1243. Qualitatively, increase in flare energy is strongly correlated with longer duration and higher amplitude. While classical flares occur near the highest energies, the majority of high-energy flares are complex in nature. The duration-energy relationship shows a difference in slope between classical (turquoise line) and complex (blue line) flares. The amplitude-duration relationship shows significantly more scatter than the other two, but still exhibits a clear trend, with longer flares generally having higher amplitude.} 
\label{fig:All_data_scatters}
\end{centering}
\end{figure*}

Figure \ref{fig:All_data_scatters} shows the relationship between flare amplitude, energy, and duration for classical and complex flares. As in Paper 1, the binning of flare durations at the low end is an artificial result of the 1-minute cadence of observations; with higher time resolution this would likely smooth out, as energy and amplitude do. This would agree with the morphological conclusions in Paper 2, which found an empirical model for classical flares that depended exclusively on relative flux and full-time at half-maximum (that is, the time between half-maximum flare-only flux during the rise and decay phases).

Following \citet{Maehara2015}, we quantify the relationship between flare duration and energy. Using our definitions of flare duration $\tau_{\mathit{Kp}}$ and energy $E_{\mathit{Kp}}$ (rather than those used by \citet{Maehara2015})\footnote{Specifically, \citet{Maehara2015} defined the duration of their ``superflares'' as the $e$-folding decay time of flare intensity after peak. Flare duration here (adopted from Papers 1 and 2) is defined as the time between the start and end of each flare, as determined by FBEYE analysis. We define $E_{\mathit{Kp}}$ separately as the luminosity of the star in the \textit{Kepler} bandpass, multiplied by its equivalent duration; we refer the reader to detailed discussion of this calculation in Paper 1 and Paper 2.}, we find that $\tau_{\mathit{Kp}} \propto E_{\mathit{Kp}}^{0.342 \pm 0.003}$ for the full set of classical flares. For complex flares, we see $\tau_{\mathit{Kp}} \propto E_{\mathit{Kp}}^{0.363 \pm 0.006}$. Both of these values are similar to, but slightly lower than, the value for ``superflares'' on solar-type stars found by \citet{Maehara2015}; however they fit well within the wide boundaries for values set by it and studies of solar flares \citep[e.g.][]{Veronig2002, Christe2008}. This suggests that, as expected, all these flares are likely caused by the same basic physical process of magnetic reconnection.

Of note is the difference in spectral index and offset between the classical and complex relationships, which indicates that complex flares have longer durations than their classical counterparts of the same energy. The complex slope and classical slope show some difference, but not a clearly significant difference between the distributions. Paper 2 found that many multiple-peak events (the so-called ``complex'' flares) were well-described by multiple overlapping classical flare templates; this potentially significant difference in the relationship between duration and energy provides further support for this assessment. Furthermore, qualitatively the complex flares exhibit somewhat lower amplitudes than classical flares of the same energy, which would also be expected from a ``superposition'' model. 

\begin{figure*}[htb]
\begin{centering}
\includegraphics[width=\textwidth]{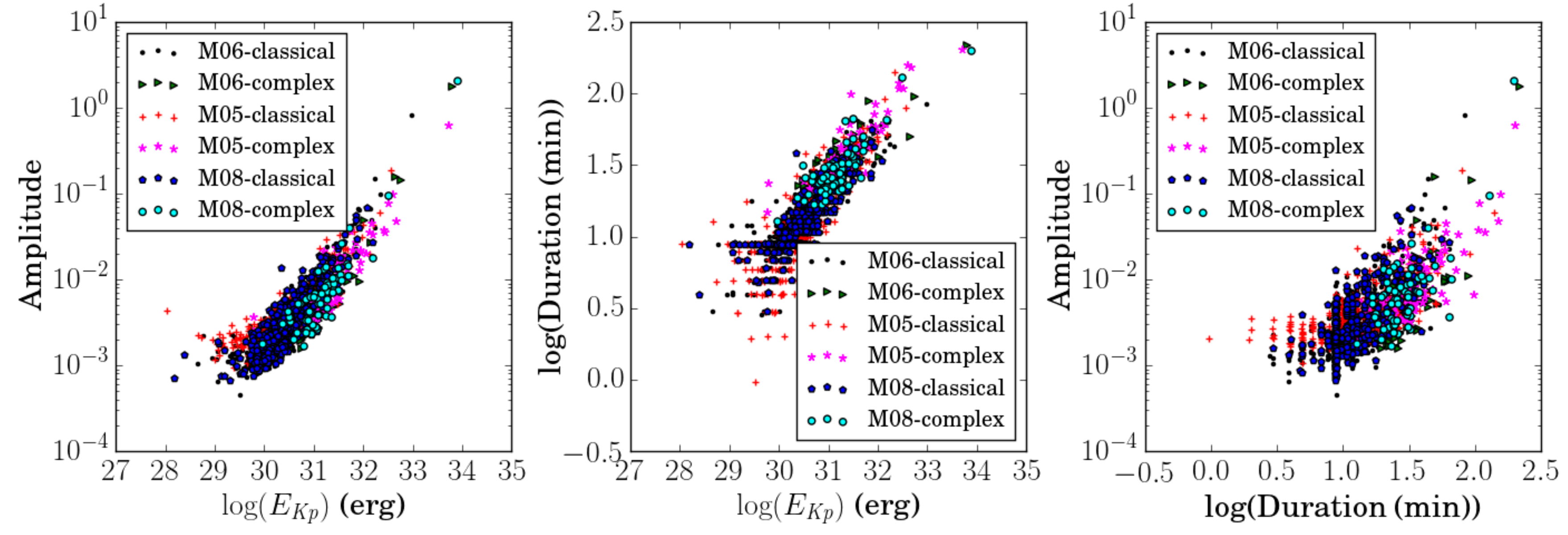}
\caption{Relationships between flare amplitude, energy, and duration for classical and complex flares for three months on GJ 1243. Qualitatively, despite having clear differences in the measured $\alpha$, all months show the same trends as the full data set, suggesting that the variation in $\alpha$ is due to the sampling of the underlying flare distribution. Note the single outlying flare at high energy in each of months 5 and 8 of the sample.} 
\label{fig:Monthly_scatters}
\end{centering}
\end{figure*}

As values of $\alpha$ show some variation over time, the question of whether the correlations here and the scatter related to them also vary merits further investigation. To examine this, with the variation in $\alpha$ in mind, we show the relationships between flare amplitude, energy, and duration for the months with the highest and lowest values of $\alpha$, and month 6, a month near the median of values, in Figure \ref{fig:Monthly_scatters}. This subset of the full data set demonstrates that for each month, the relationships between these three variables are qualitatively identical to the full data set, for both classical and complex flares. Furthermore, while the overall trend appears to be the same month-to-month, months 5 (lowest $\alpha$) and 8 (highest $\alpha$) demonstrate a clear gap, much larger than that observed in the full data set, between the highest-energy and next-highest-energy flares in all three attributes. This suggests that the values for $\alpha$ in these time segments (as well as the large range of possible values for $\alpha$ seen in Figure \ref{fig:alph_v_time}) are due to these outliers in how the underlying distribution was sampled.

\begin{figure*}[htb]
\begin{centering}
\includegraphics[width=\textwidth]{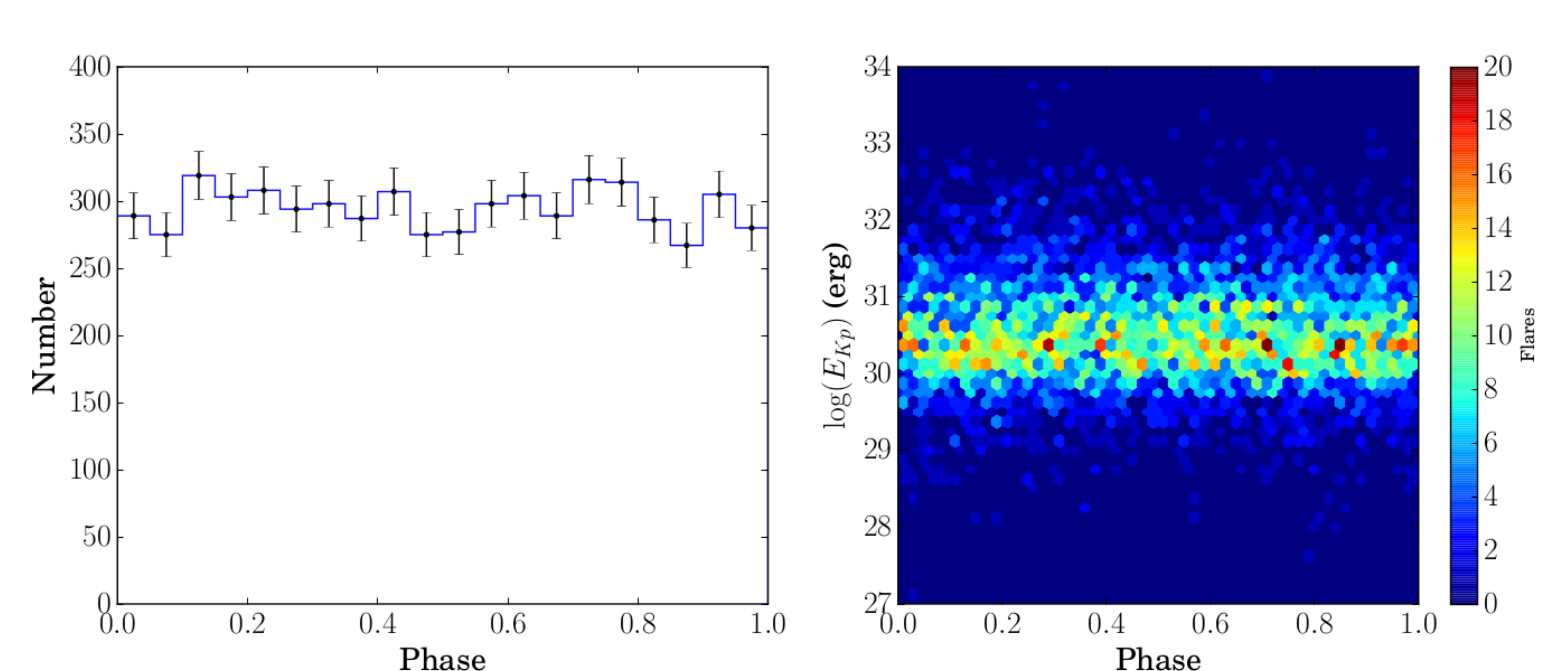}
\caption{Flare occurrence and energy as a function of stellar phase. Left panel illustrates that for the full eleven-month period, the number of flares in each phase bin remains very flat, consistent with no trend. Right panel shows a similar lack of corellation of flare energy with phase.} 
\label{fig:phaseplots}
\end{centering}
\end{figure*}

\subsection{Flare Timing with Respect to Starspot Modulation}

Long- and short-cadence \textit{Kepler} observations of GJ 1243 show a regular, $\sim 0.59$ day periodicity in its brightness, believed to be due to a persistent starspot or region of starspots \citep{Davenport2015}. As follow-up to the analysis of correlation with starspot modulation in Paper 1, we also examined the distribution of flares as a function of star phase month-by-month as well as for the full eleven-month sample, to observe whether flares were more likely to occur when the largest fraction of the visible hemisphere was covered with starspots. Figure \ref{fig:phaseplots} shows the histogram for the full data set and demonstrates no clear trend, as was found in Paper 1. Similar results are seen for the individual months and quarters. Similarly, there is no apparent trend in flare energy as a function of phase, the same result as Paper 1.

\citet{Davenport2015} identified a long-lived starspot at higher latitude on GJ 1243 (possibly a spot cap or group) which holds very stable in phase (standard deviation in longitude of only 16 degrees over four years of data). They were able to weakly constrain the latitude of the ``primary starspot'' to a region that implies that the spot covers a significant fraction of the pole. Given the inclination derived in Section \ref{section:kinematics}, part of the spot will always be in view, leading to a minimal correlation between flare occurrence and phase even if the flares are associated with the spot.

\section{Behavior of Flares in Spectra}
\label{section:spectra}

\begin{figure}[htb]
\begin{centering}
\includegraphics[width=0.45\textwidth]{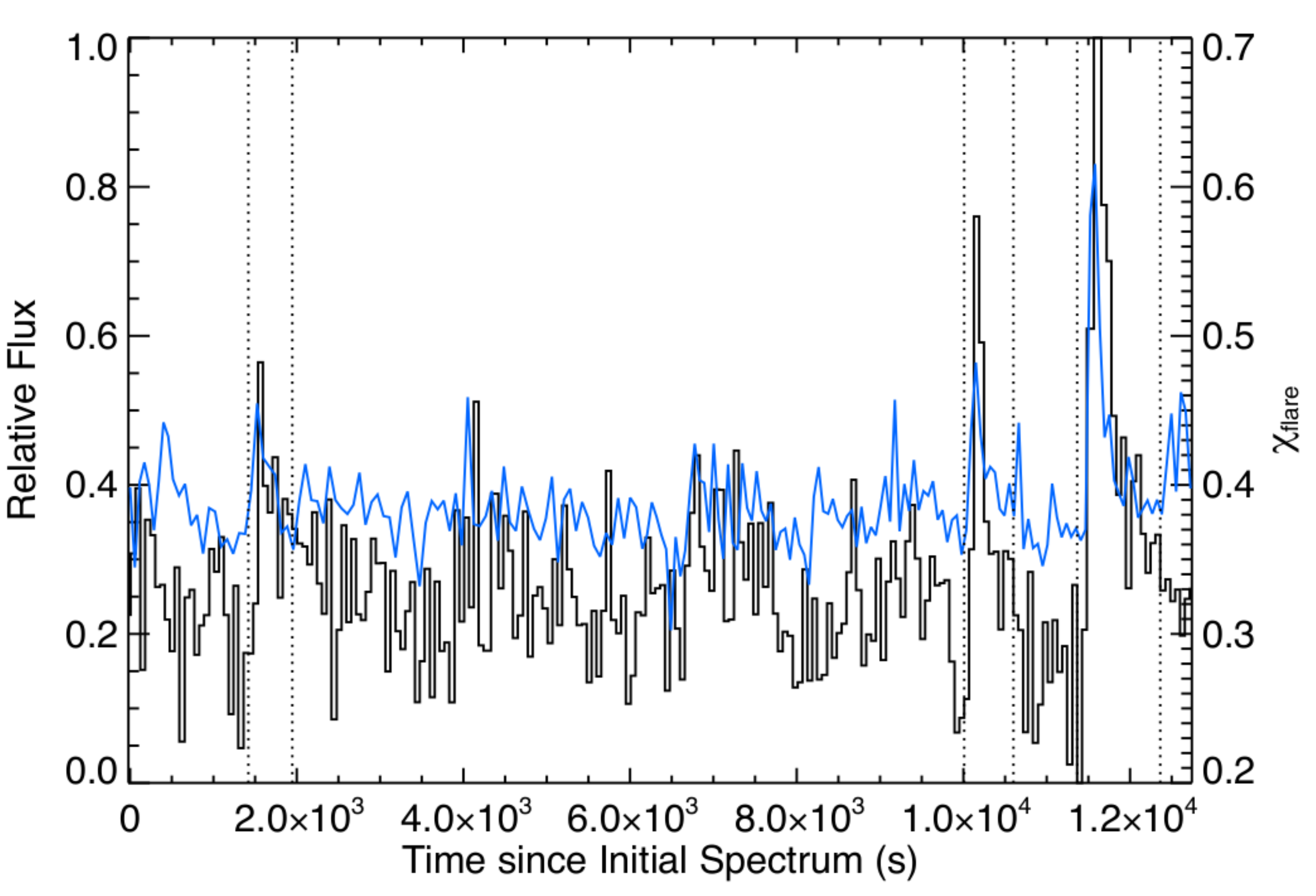}
\caption{The period of simultaneous \textit{Kepler} and spectroscopic observation. The black histogram (left $y$-axis) is the observed relative flux from \textit{Kepler}, plotted against time from the midpoint of the first spectrum recorded. The blue line (right $y$-axis) is the observed $\chi_{\mathrm{flare}}$ (see Section 5.1) for each full (quiescent + flare) spectrum. The vertical dotted line indicates flare start and end times.}
\label{fig:chiflare_v_kepler}
\end{centering}
\end{figure}

We recorded simultaneous spectroscopy and \textit{Kepler} photometry of three low-energy classical flares, as shown in Figure \ref{fig:chiflare_v_kepler}; the three flares have \textit{Kepler} energies $\log(E_{\mathit{Kp}}) \sim 30.1-30.7$, near the low-energy break of the cumulative FFD from a power law seen in Figure \ref{fig:All_data_FFD}. These energies are quite low in comparison to those analyzed in \citet{Kowalski2013}. Following the adopted conversion from $E_{\mathit{Kp}}$ to $E_{U}$ from Paper 1, $E_{U}$ here ranges from $7 \times 10^{29}$ ergs to $3 \times 10^{30}$ ergs, the latter of which is lower in energy than the lowest-energy flare in \citet{Kowalski2013}. Because of this, spectral data for these flares have a higher relative noise threshold than the flares analyzed in \citet{Kowalski2013}. We present light curves of the normalized spectral and \textit{Kepler} quantities as a function of time elapsed from flare start. We focus our analysis of these ``spectroscopic flares'' on the light curve for the final of the three flares observed, as its higher energy produced by far the clearest differentiation from the quiescent pre-flare spectra.

To mitigate issues due to the residual airmass effect discussed in Section 2, we choose a quiescent spectrum just prior to the start of each flare for use in isolating the flare-only emission, rather than generating a master quiescent spectrum from the entire night. With this amendment, flare-only spectra were generated following the procedure outlined in \citet{Kowalski2013}. A standard quiescent or pre-flare spectrum was selected for each flare. A scaling was then applied to each flare spectrum in a given flare relative to the adopted standard quiescent spectrum. Each spectrum during the flare was multipled by a range of possible scale factors (0.8-1.2), from which we subtracted the standard quiescent spectrum. We then calculated the sum of the standard deviation of the subtraction residuals in three spectral regions, avoiding atmospheric features, from $\lambda = 6600-6800$ \AA, $\lambda = 7000-7100$ \AA , and $\lambda = 7350-7550$ \AA. These regions correspond to molecular bandheads which produce strong flux changes in the quiescent spectrum; errors in flux scaling appear as significant over- or under-subtractions at these wavelengths. We adopted the scale factor which minimized the sum of the standard deviations as the best scale factor for each flare spectrum, and then applied this scaling factor to the total (quiescent + flare) spectrum, before subtracting the quiescent values to produce the flare-only spectrum.

\subsection{Spectral Observational Parameters}

\citet{Kowalski2013} defined several observational parameters for spectra, which we adopt here. Briefly, we use the average flux measurements C3615 and C4170, their ratio $\chi_{\mathrm{flare}}$, and the flux measures BaC3615, $\mathrm{BaC_{tot}}$, and PseudoC. We refer the reader to \citet{Kowalski2013} for definitions of these measures. In addition to these, we record line fluxes for the Balmer series through H$\delta$, as well Ca II K, using the same line and continuum windows used in \citet{Kowalski2013} for consistency.

All times recorded for both ground- and space-based observations are times at the midpoint of the observation. All times from the spectroscopy (generally given in MJD) have been converted to $\mathrm{BJD}-2454833$, a standard time for the \textit{Kepler} data which was used in Papers 1 and 2, to ensure synchronization. 



\subsection{The Third Flare}
\label{sec:flarethree}

\begin{figure*}[htb]
\plottwo{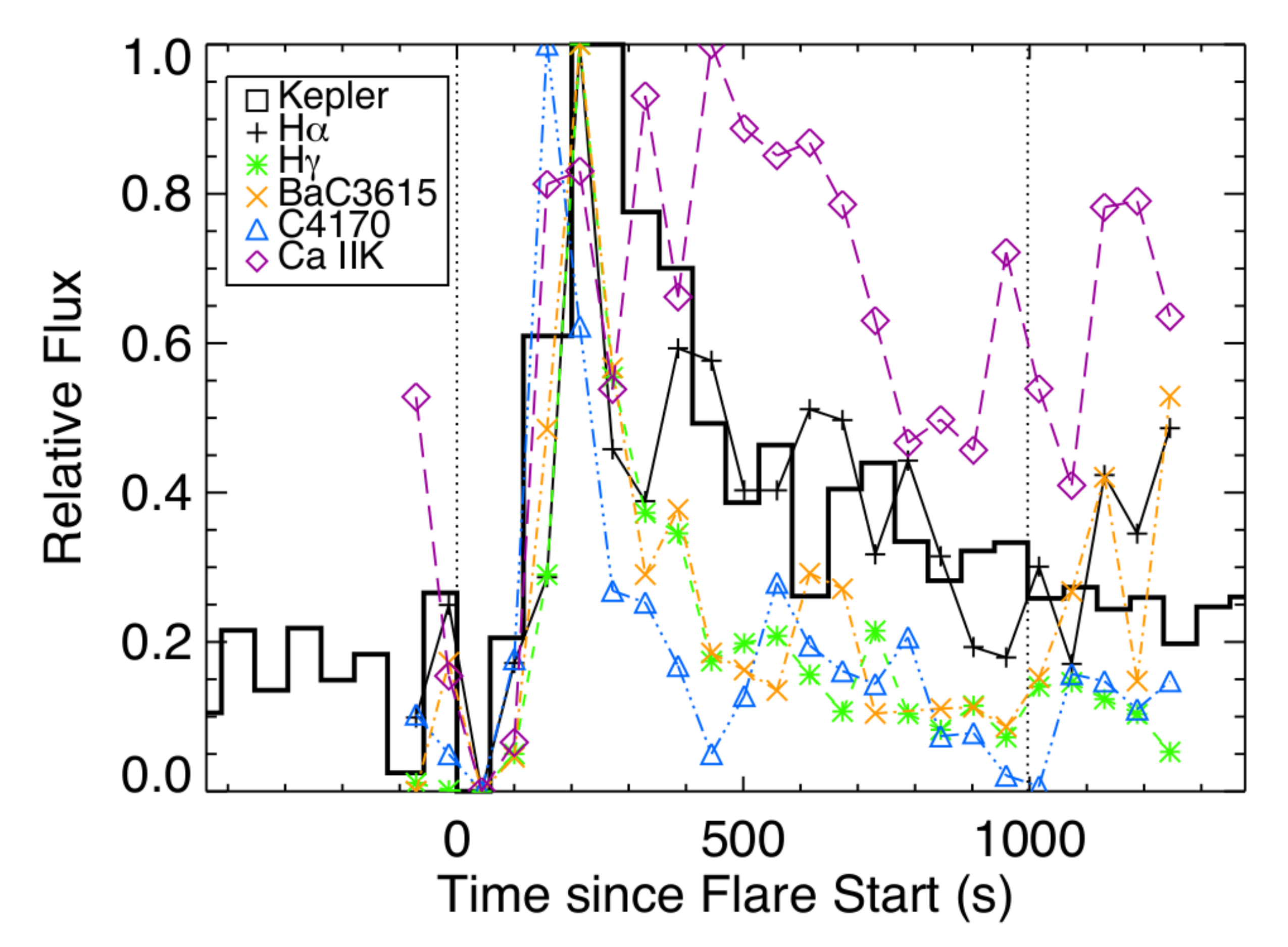}{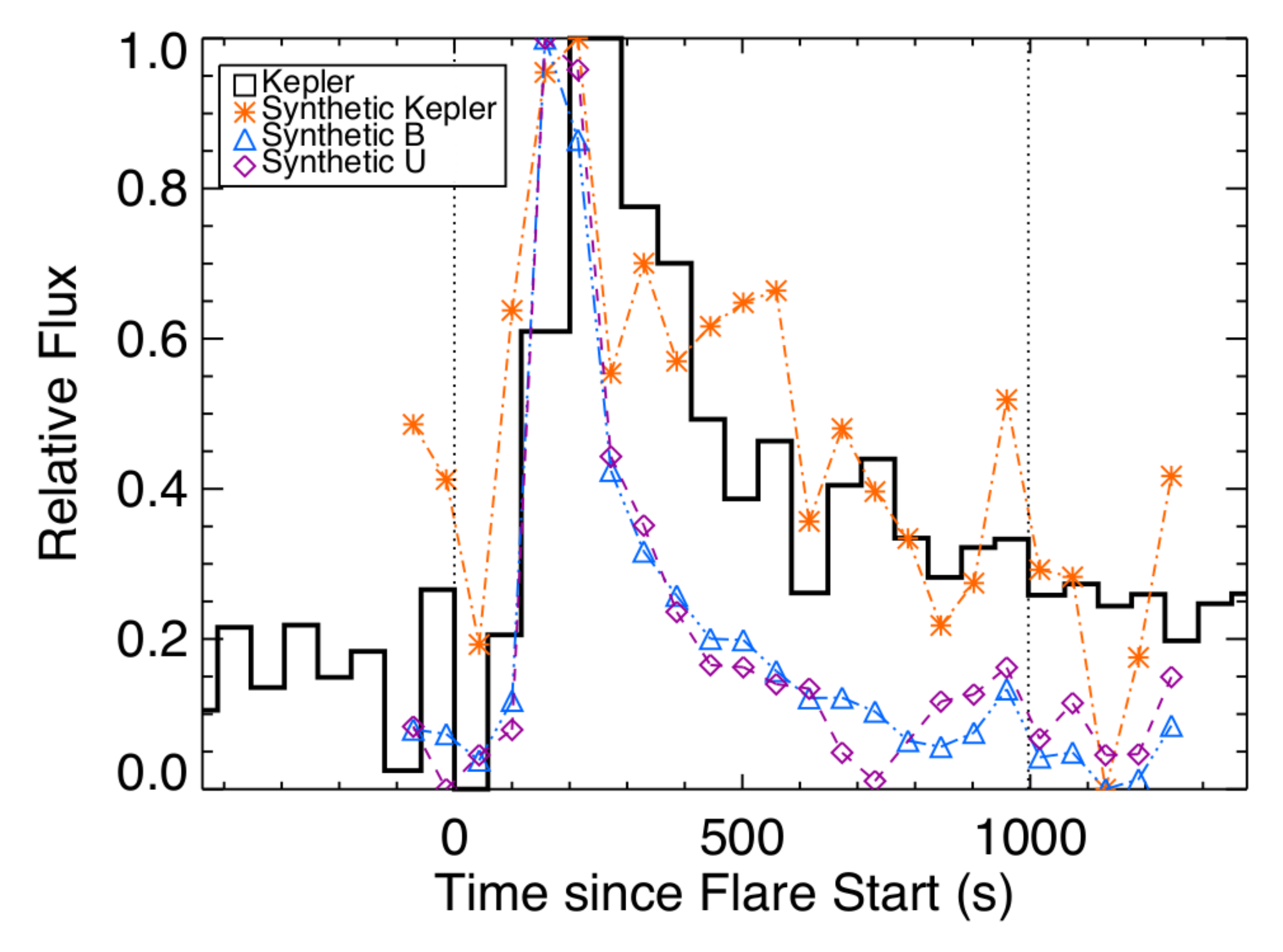}
\caption{Light curves for the third spectrosopic flare. \textbf{a}. Line and continuum evolution from DIS spectra. H$\beta$ and H$\delta$ evolution (not plotted) follows the evolution of H$\gamma$. PseudoC evolution (not plotted) follows the evolution of BaC3615. Errors for HB lines are generally $\sim 0.1$ in relative flux space. \textbf{b}. Relative \textit{Kepler} fluxes, and synthetic $U$, $B$, and \textit{Kepler} fluxes from DIS spectra.}
\label{fig:Flare_3_plots}
\end{figure*}

The third flare began at \textit{Kepler} date 1232.9472656 and lasted for 15.64 minutes. The recorded $\log E_{\mathit{Kp}}$ of the flare was $30.66$, nearly four times as strong as the first flare. As seen in Figure \ref{fig:Flare_3_plots}a, this flare exhibits a delay of one minute between maximum C4170 and maximum Balmer emission. While there is a delay, the minimal size (the smallest detectable delay in our time resolution) suggests that these peaks were produced by a common heating mechanism, as with most classical impulsive events examined in \citet{Kowalski2013}.

\subsection{Flare Morphology}


The shape of the flares with simultaneous photometry and spectroscopy suggests that they are ``low-amplitude,'' ``classical'' flares, using broad qualitative descriptors from \citet{Kowalski2013}. \citet{Kowalski2013} defined a quantitative ``impulsiveness index'' $\mathcal{I}$ based on the full width of the $U$-band light curve at half maximum and the quantity $I_{f,U}$ defined in \citet{Gershberg1972}. A similar impulsiveness index could be defined for the \textit{Kepler} data, but would yield very little information without multiple flares with differing levels of impulsiveness. Instead, we calculate a rough filter-weighted count flux measurement for flare 3 in the $U$-band, based on our spectral data. To avoid extreme noise effects at the blue end of the spectrum and the portion of the bandpass missing from our spectra, we set the filter sensitivity to 0 blueward of 3400 \AA. The results (along with synthetic $B$-band flux, actual \textit{Kepler} flux, and a synthetic measurement of \textit{Kepler} flux from spectra) are shown in Figure \ref{fig:Flare_3_plots}. Following Section 2.1 from \citet{Kowalski2013} to find values $I_{f, U}$ and $t_{1/2, U}$ for these data, we calculate $\mathcal{I}_{U} \sim 0.4$ for this flare. To confirm this approximation of $\mathcal{I}_{U}$, we look to other spectral diagnostics. \citet{Kowalski2013} showed an inverse relationship between impulsiveness of flare and ratio of BaC emission to total emission blueward of the Balmer jump (as measured by BaC3615/C3615); ``gradual flare'' events had much higher percentages ($55-80\%$) of C3615 due to BaC3615 at the peak than ``impulsive'' or ``hybrid'' events. All three ``spectroscopic'' flares exhibited BaC3615/C3615 ratios of $\sim 0.7-0.81$ at peak (Table \ref{table:HB_data}), indicating that all were ``gradual flare'' events. The relationship between the $U$, $B$, and \textit{Kepler} bands is discussed further in Section \ref{sec:UBKeprel}.




\subsection{Emission Lines and the Hydrogen Balmer Flux Budget}

\begin{deluxetable*}{lccccccc}
\tablecaption{Hydrogen Balmer (HB) Properties
\label{table:HB_data}}
\tablecolumns{8}
\tablehead{\colhead{Flare ID} & 
    \multicolumn{2}{c}{HB flux/Total flux} & 
    \colhead{H$\delta$/Total flux} & 
    \colhead{H$\gamma$/Total flux} & 
    \colhead{H$\beta$/Total flux} & 
    \multicolumn{2}{c}{BaC3615/C3615} \\
    \colhead{} & 
    \colhead{Peak} & 
    \colhead{Decay} & 
    \colhead{Peak} & 
    \colhead{Peak} & 
    \colhead{Peak} & 
    \colhead{Peak} & 
    \colhead{Decay} }
    
\startdata
Flare 1 & 0.5  & \nodata & 0.04 & 0.05 & 0.05 & 0.8  & \nodata \\
Flare 2 & 0.4  & 0.3  & 0.03 & 0.02 & 0.03 & 0.7  & 0.6 \\
Flare 3 & 0.45  & 0.7  & 0.051 & 0.049 & 0.057 & 0.81  & 1.2 \\
\enddata

\end{deluxetable*}

To constrain unexplained energy sources for model predictions (e.g. white-light continuum emission), we measure flux attributable to hydrogen Balmer (HB) emission relative to the total flare emission. We quantify this by summing the fluxes in H$\delta$, H$\gamma$, H$\beta$, PseudoC, and $\mathrm{BaC_{tot}}$. As in \citet{Kowalski2013}, we measure the percentage of HB flux to total flux in spectra blueward of 5200 \AA\ (\%HB) at peak emission and the beginning of the gradual decay phase. These data provide direct comparison to RHD models. Data for each flare is listed in Table \ref{table:HB_data}; here, we summarize some key findings for the third flare.

Like \citet{Kowalski2013}, we find that in the third flare, the fraction of emission attributable to Balmer emission increases from peak continuum emission to the beginning of the gradual decay phase by $\sim 20\%$. However, we also observe a dramatic ($\sim 30\%$) \textit{decrease} in \%HB immediately afterward. Additionally, the third flare has $50 - 60\%$ of total flux from HB emission at the beginning of the gradual decay phase, which also corresponds well to GF-type characteristics. At peak, each of the H$\beta$, H$\gamma$, and H$\delta$ lines produces $\sim 5\%$ of the emission, in line with \citet{Kowalski2013}.


\subsection{Spectral Implications for the Empirical Flare Model}

Both Paper 2 and \citet{Kowalski2013} found specific changes in flare behavior between the impulsive and gradual decay phases. Paper 2 found this change in its two-component flare decay model, while \citet{Kowalski2013} found that the gradual decay phase was marked by an increase in hydrogen Balmer flux (as a percentage of total flare flux) of $\sim 20\%-30\%$. Here, we use the results of spectral analysis of the third flare (Figure \ref{fig:Flare_3_plots}a) to explain the empirical model from Paper 2. The \textit{Kepler} response function has minimal transmission below 4000 \AA\footnote{Kepler Instrumentation Handbook, section 3.3, document number KSCI-19033-001 of 2009 July 15.}, so NUV components of the flare produce a minimal contribution to the overall shape. However, the lower-order Balmer lines (H$\alpha$, H$\beta$, H$\gamma$, and H$\delta$) contribute to the \textit{Kepler} flux, as does C4170.

The impulsive rise is most characterized by continuum emission (as represented by C4170), as expected. C4170 rapidly increases from its quiescent pre-flare level; the total (quiescent + flare) value of C4170 at peak is 1.15 times greater than the total value immediately pre-flare. C4170 drops back to near-quiescent levels fairly quickly (within five spectra of its peak), corresponding with the flare's impulsive decay phase, and then evolves similarly to the evolution of the \textit{Kepler} light curve, gradually decaying until the end of the flare. This agrees well with the canonical $\sim 10^{4}$ K ``blackbody'' emission component, as expected. The HB lines begin at a lower quiescent point than C4170 does, and peak one spectrum after C4170 does, in line with the \textit{Kepler} peak. To evaluate the increase in line emission, we calculate the ratio of emission from each line to H$\gamma$ in total (quiescent + flare) spectra; to evaluate H$\gamma$, we calculate the ratio of H$\gamma$/C4170. At peak, values of the lines range from 2.1 (H$\gamma$/C4170) to 7.1 (H$\delta$/H$\gamma$) times their quiescent levels. Like C4170, these four lines show a rapid decay. The bluer lines (all but H$\alpha$) then exhibit a more gradual fade at approximately the same point that C4170 enters its gradual decay (approximately corresponding to the start of the Paper 2 gradual decay phase), while H$\alpha$ mimics the \textit{Kepler} light curve.

Paper 2 characterized their two-component empirical decay as representing two physically distinct regions, with independent cooling profiles, producing radiation throughout the flare decay. Our analysis of the third flare here suggests that the secondary component of this model connects the gradual decay phase blackbody (represented by C4170) and the increase in relative proportion of HB flux (Table \ref{table:HB_data}), as was found in \citet{Kowalski2013}. This implies that the blackbody and HB components of the gradual decay phase are physically connected to each other, originating from the same physical process, but are \textit{physically distinct} from the impulsive phase of the flare, either by virtue of different spatial region, different atmospheric layer, or different cooling process. This result is critical for future modeling, as it suggests that the gradual decay phase and impulsive phase may potentially be modeled separately, but with a connected underlying physical origin. Current modeling efforts \citep[e.g.][]{Kowalski2015} focus on replicating the impulsive phase with 1-dimensional RHD multi-thread modeling \citep[e.g.][]{Warren2006}; the gradual phase currently remains a separate challenge, due to the quite different timescales. Given the distinct characteristics of their evolution, it is unlikely that a single underlying fundamental heating/cooling mechanism can produce both the observed impulsive and gradual phases, unless the impulsive phase drastically alters the initial conditions of the flaring region for the gradual phase. As such, modeling the gradual phase as a separate, temporally-overlapping problem will allow for more focused efforts to improve both models. For example, there is no known explanation for the observed evolution of Ca IIK, which peaks during the transition from impulsive to gradual decay for the third flare, though that evolution does indicate a less intense heating/cooling mechanism than the impulsive phase's electron-beam heating. A separate gradual-phase model with an independent heating and cooling mechanism would be able to directly address this, as well as the observed blackbody and HB evolution.

\subsection{Extension to Other Stars}
\label{sec:UBKeprel}


Connecting the results of studies of stellar flares from \textit{Kepler} data to the literature is challenging, as most ground-based flare studies utilize $U$ and $B$ band optical data \citep[see, e.g.][]{LME76, Kowalski2013}, while \textit{Kepler} data lack color information. To enhance comparison of \textit{Kepler} flare data to canonical studies, it will be critical to find an appropriate conversion of measurements of $L_{\mathit{fl}}$, the flare luminosity in a given bandpass, to $L_{\mathit{fl}}/L_{\mathrm{bol}}$, the bolometric luminosity for the star.

As a preliminary step in this direction, we analyze the relationship between synthetic $U$, synthetic $B$, and synthetic and actual \textit{Kepler} data, presented for flare 3 in Figure \ref{fig:Flare_3_plots}. Qualitatively, the synthetic \textit{Kepler} data provides a satisfactory match to the observed data from \textit{Kepler}, but with some additional noise due to observed fluctuations in the continuum level of the red end of the spectra.  We therefore determine relationships between relative $U$, $B$, and \textit{Kepler} fluxes based on these synthetic fluxes rather than the actual \textit{Kepler} fluxes, as the spectral fluxes are synchronized.

\begin{figure}[htb]
\begin{centering}
\includegraphics[width=0.45\textwidth]{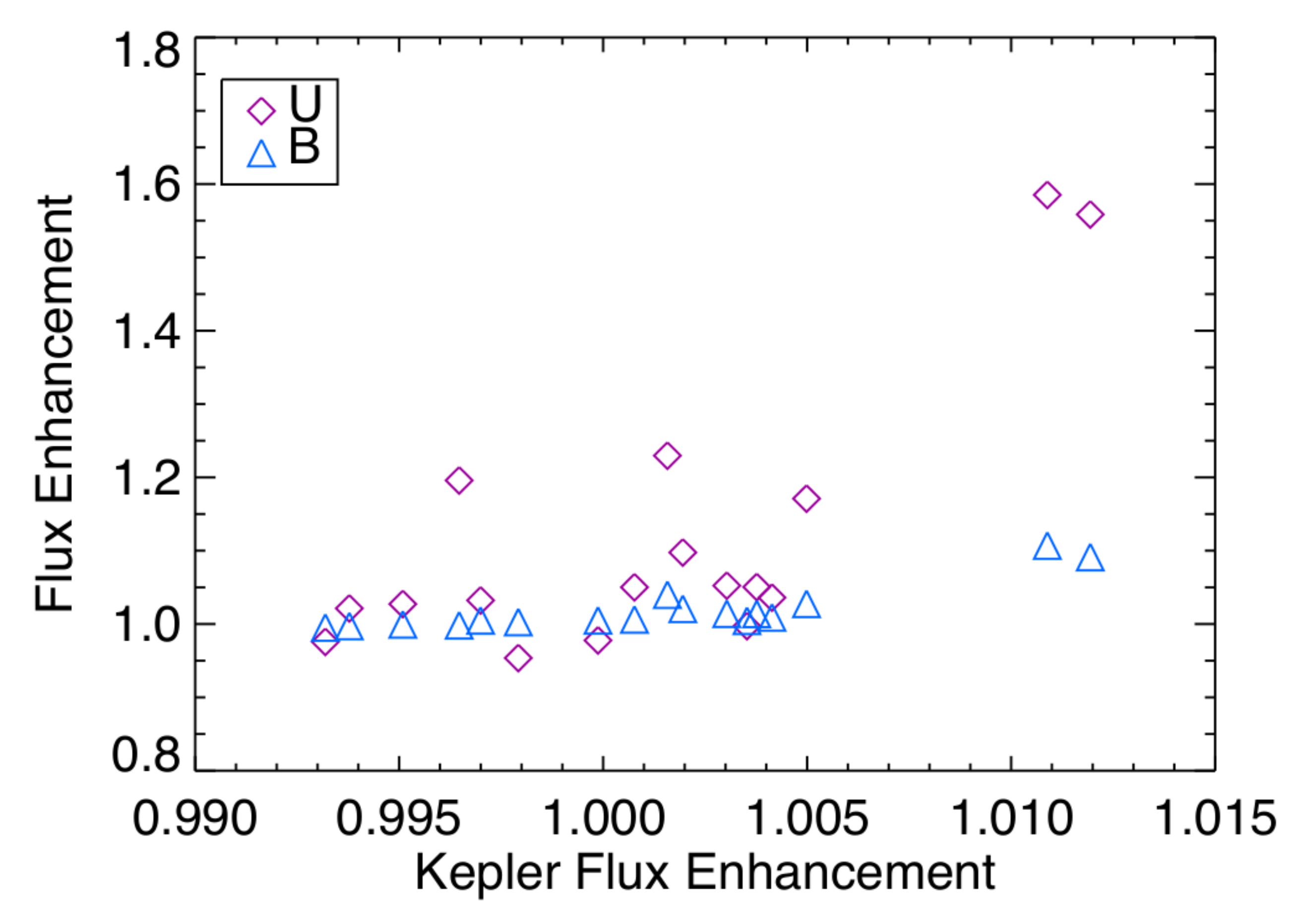}
\caption{Flux enhancement $\Delta(f)_{\lambda}$ of synthetic $U$ and $B$ fluxes for the third flare as a function of synthetic \textit{Kepler} enhancement for the third flare. Flux enhancement, similar to $I_{f} + 1$ from \citet{Gershberg1972}, measures the ratio of full (quiescent + flare) light to quiescent light.}
\label{fig:synthscat}
\end{centering}
\end{figure}

We define $\Delta(f)_{\lambda}$ as the flux enhancement during a flare for each band. This is analogous to the $I_{f} + 1$ quantity from \citet{Gershberg1972}, but utilizes the calibrated flux rather than the count flux. Figure \ref{fig:synthscat} presents the flux enhancements $\Delta(f)_{\lambda}$ for the $U$ and $B$ bands for the third flare as a function of the \textit{Kepler} flux enhancement. The synthetic \textit{Kepler} flux values are only enhanced significantly ($> 3 \sigma$) above the quiescent noise level during the two observations at the peak of this flare; as such, we focus on flux enhancement conversions during the peak.

We find that for the peak of the third flare,

\begin{equation}
\Delta(f)_{B}, \mathrm{peak} = 1.09 (\pm 0.01) \Delta(f)_{\mathit{Kep}, \mathrm{peak}}
\end{equation}

\noindent
and

\begin{equation}
\Delta(f)_{U, \mathrm{peak}} = 1.55 (\pm 0.02) \Delta(f)_{\mathit{Kep}, \mathrm{peak}},
\end{equation}

\noindent
suggesting (as expected) that a small flux enhancement in the \textit{Kepler} bandpass indicates a similarly-sized enhancement in $B$, and a much larger enhancement in $U$. Future work (J. R. A. Davenport et al., in prep) will expand on this exploration of conversion from the \textit{Kepler} bandpass to the conventional $U$ and $B$ filters, using the flare light curve template for GJ 1243 from Paper 2 and robustly-determined bolometric flare characteristics.

\section{Discussion and Conclusions}
\label{sec:Conclusions}

In this paper, we have presented a comprehensive study of the star GJ 1243, using previous data in the literature, 11 months of short-cadence wide-band photometry from \textit{Kepler}, and simultaneous spectroscopy for three flares. We calculated and analyzed the cumulative frequency of flares as a function of energy; the relationships between flare energy, amplitude, and duration; the relationship of flare frequency and energy with stellar phase; and several characteristics of flares based on spectroscopy. In the following sections, we discuss the potential for a secondary power law for frequency of the highest-energy flares, potential origins of the flare morphology observed in the \textit{Kepler} data based on the spectral data, and potential extensions of the discoveries made here to other stars.

\subsection{Basic Characteristics of the Star}

We demonstrated based on high-resolution echelle spectroscopy that GJ 1243 has a large $v \sin i$ of 25 $\mathrm{km s^{-1}}$, as expected for its high activity level. We used the observed period in its \textit{Kepler} light-curve and this measurement to find that the system's inclination angle $i$ was $32^{\circ}$. Using data on its proper motion, distance, and radial velocity, we showed that GJ 1243 is likely (98.5\% probability) a member of the Argus association, suggesting that its age is 30-50 Myr. This relatively young age agrees well with the high level of activity we observe on this star.

\subsection{Frequency of High-Energy Flares}
We find that generally a single power-law model fits each data set in monthly and quarterly increments, as well as the full data set. Over time, we show that $\alpha$ shows some uncorrelated variability over time, likely due to the varying sample of flares over each epoch studied. Interestingly, the values of $\alpha$ for each subset are generally lower than the value for the full dataset. We believe this is in part due to the sample size of the full data set allowing us to probe the FFD at a higher energy resolution, compared to the subsets.

Additionally, we find that the highest energy flares, both saturated and unsaturated, deviate from a single power law in a way that suggests a steeper power law than the full sample. This would have significant implications for stellar modeling, as values of $\alpha > 2$ are a possible mechanism for the heating of the solar corona. However, these heating models require this larger value of $\alpha$ to extend to low energies, which our data do not. For the full data set $\alpha \approx 2$, though there is a turn-off at low energies where flares above the minimum energy threshold are not detected as expected, as was observed in Paper 1. Paper 2 showed that the 11-month sample used here is 90\% complete at the highest energy levels, suggesting that there may be ``missing'' flares from the sample that could account for this. Additionally, some lower-energy ``classical'' events could have been incorporated into ``complex'' events. While Paper 2 found that this potential overlap could not account for all of the ``missing'' portion, it could be an additional contributing factor.

\subsection{Correlations of Flare Characteristics}

We find a strong, positive correlation between flare energy and flare duration, as well as between flare energy and amplitude, similar to the trends found in Paper 1. We find that complex flares have higher energies and longer durations than classical flares at the same amplitude. We quantify the relationship between duration and energy, and find that complex and classical flares show a distinct difference in this relationship, which agrees with the characterization of some complex flares as a superposition of classical flares, presented in Paper 2. As with Paper 1, we find no correlation of flare timing or energy with the phase caused by the persistent starspots on the surface, in monthly or quarterly time periods or in the full data set. This suggests that flaring is distributed across the stellar surface, rather than concentrated only in the regions where these starspots appear. Additionally, the high-latitude constraint for the persistent spot found in \citet{Davenport2015}, combined with the inclination angle found in Section 3, indicates that there would be minimal correlation between flare occurrence and phase were the flares in fact associated with the spot.

\subsection{Flare Spectroscopy}

We observed three low-energy flares with simultaneous ground-based spectroscopy and \textit{Kepler} photometry. Using this simultaneous photometric and spectroscopic data, we were able to classify the third flare of this set as a ``gradual-flare'' type event, following the definition from \citet{Kowalski2013}. We were also able to derive an impulsiveness index $\mathcal{I}$ for this flare based on the \textit{Kepler} data, finding that this value was two orders of magnitude lower than the same index in the $U$ band (synthesized from spectra), due to the smaller flare-induced flux enhancement in the \textit{Kepler} bandpass. We also presented a spectroscopy-based explanation for the two-component empirical flare model adopted in Paper 2, finding that the gradual phase HB and blackbody were tied to each other, but resulted from a cooling region physically distinct from that of the impulsive phase. This suggests that the impulsive and gradual phases are physically distinct phenomena potentially arising from the same initial conditions, which allows for the possibility of separately modeling the impulsive and gradual phases of the flare.

\subsection{Application to Future Flare Studies}

Using simultaneous photometric and spectroscopic data, we were able to develop a relationship between flux recorded in the \textit{Kepler} bandpass and in the more traditional $U$ and $B$ bandpasses at the peak of a faint flare, directly connecting this result to the larger canon of previous flare surveys. This basic conversion will be invaluable for future studies of dMe flares recorded with \textit{Kepler}, and will be improved by future work that will approach the problem beginning with the empirical \textit{Kepler} flare model.


\acknowledgements

We thank the referee for providing feedback that improved the content and clarity of this paper. We gratefully acknowledge support for this work from NASA \textit{Kepler} Cycle 2 GO grant NNX11AB71G, NASA \textit{Kepler} Cycle 3 GO grant NNX12AC79G. A.F.K. acknowledges the support from NSF grant AST08-07205, the NASA Postdoctoral Program at the Goddard Space Flight Center, adminstered by Oak Ridge Associated Universities through a contract with NASA, and from UMCP GPHI Task 132. J.R.A.D. is supported by an NSF Astronomy and Astrophysics Postdoctoral Fellowship under award AST-1501418. S.M.S. wishes to thank Evan A. Rich and Michael Malatesta for valuable discussion of spectral reduction and analysis, and Jonathan Gagn\'e for valuable insights into BANYAN II.

Observations reported here were obtained with the Apache Point Observatory 3.5m telescope, which is owned and operated by the Astrophysical Research Consortium.	

This paper includes data collected by the \textit{Kepler} mission. Funding for the \textit{Kepler} mission is provided by the NASA Science Mission directorate. Some of the data presented in this paper were obtained from the Mikulski Archive for Space Telescopes (MAST). STScI is operated by the Association of Universities for Research in Astronomy, Inc., under NASA contract NAS5-26555. Support for MAST for non-\textit{Hubble Space Telescope} data is provided by the NASA Office of Space Science via grant NNX13AC07G and by other grants and contracts.

\end{document}